\documentclass[a4paper]{JHEP3}
\pdfoutput=1

\usepackage[centertags]{amsmath}
\usepackage{amssymb,amstext,amsmath}
\usepackage{graphicx}
\usepackage{epsfig}
\usepackage[stable]{footmisc}
\usepackage{appendix}

\usepackage{young}
\usepackage{pbox}

\newcommand{\be}{\begin{equation}}
\newcommand{\ee}{\end{equation}}

\preprint{WIS/12/15-NOV-DPPA}

\title{Baryons, monopoles and dualities in Chern-Simons-matter theories}

\author{
Ofer Aharony \\
\it{Department of Particle Physics and Astrophysics,\\
Weizmann Institute of Science, Rehovot 7610001, Israel}\\
E-mail : {\tt Ofer.Aharony@weizmann.ac.il}
}

\abstract{There is significant evidence for a duality between (non-supersymmetric) $U(N)$ Chern-Simons
theories at level $k$ coupled to fermions, and $U(k)$ Chern-Simons theories at
level $N$ coupled to scalars. Most of the evidence comes from the large $N$ 't Hooft limit,
where many details of the duality (such as whether the gauge group is $U(N)$ or
$SU(N)$, the precise level of the $U(1)$ factor, and order one shifts in the level) are not important. 
The main evidence for the validity of the duality at finite $N$ comes from adding masses
and flowing to pure Chern-Simons theories related by level-rank duality, and from flowing
to the non-supersymmetric duality from supersymmetric dualities, whose finite $N$
validity is well-established. In this note we clarify the implications of these flows for
the precise form of the duality; in particular we argue that in its simplest form the
duality maps $SU(N)$ theories to $U(k)$ theories, though there is also another
version relating $U(N)$ to $U(k)$. This precise form strongly affects the mapping
under the duality of baryon and monopole operators, and we show, following
arguments by Radi\v cevi\'c, that their mapping is consistent with our claims. We also
discuss the implications of our results for the additional duality between these Chern-Simons matter
theories and (the UV completion of) high-spin gravity theories on $AdS_4$. The latter theories should contain heavy
particles carrying electric and/or magnetic charges under their $U(1)$ gauge symmetry.
}

\begin{document}

\section{Introduction}

One of the main tools to study strongly coupled field theories is duality -- the claim that certain
strongly coupled theories are equivalent to (generally different) weakly coupled theories. This
phenomenon was first discovered in two space-time dimensions, and in the past twenty years
many examples of duality were discovered in supersymmetric theories in higher dimensions.
However, the theories which we would most like to understand at strong coupling are
non-supersymmetric theories in three and four space-time dimensions. For such theories
there is only one conjectured duality so far -- the duality \cite{Giombi:2011kc,Aharony:2011jz,Aharony:2012nh} between three dimensional
$U(N)$ Chern-Simons (CS) theories coupled to fermions, and three dimensional 
$U({\tilde N})$ Chern-Simons theories
coupled to scalars (there is a similar duality also for $SO(N)$ theories, that we will
not discuss here). 

This non-supersymmetric duality was originally motivated by the duality \cite{Klebanov:2002ja,Sezgin:2003pt} of both types
of theories, in the large $N$ limit,
to high-spin gravity theories on $AdS_4$ constructed by
Vasiliev \cite{Vasiliev:1992av}. It was then corroborated by many exact computations in the large $N$
limit  (with fixed $N/k$, where $k$ is the Chern-Simons level), which are all consistent with the duality, and by the classification of theories with slightly broken
high-spin symmetries \cite{Maldacena:2011jn,Maldacena:2012sf}. It was also generalized \cite{Chang:2012kt,Jain:2013gza} to theories with
additional fields, including supersymmetric field theories for which some of the dualities
that arise were already previously known \cite{Giveon:2008zn,Benini:2011mf,Aharony:2013dha,Aharony:2014uya}.

All known dualities that are valid at large $N$ also have a generalization to finite $N$,
so it is natural to conjecture that this is true also here. However, it is difficult to test this
conjecture, since it is not known how to perform the large $N$ computations at finite
$N$ (and even the $1/N$ corrections have not been computed yet). In particular,
the large $N$ tests do not distinguish between theories whose level is shifted by a
finite number in the large $N$ limit, nor between $U(N)$ and $SU(N)$ theories.
So far there are
just two pieces of evidence for the finite $N$ duality. One piece of evidence comes
from deforming the theories related by the duality by a fermion mass term \cite{Aharony:2012nh}. At low
energies the CS-fermion theory flows to a pure Chern-Simons theory, with a level
depending on the sign of the mass term. Similarly, the CS-scalar theory flows
to a pure Chern-Simons theory, with the rank depending on the sign of the mass term.
The resulting Chern-Simons theories must then be equivalent, and indeed, for a specific
choice of the order one shift in the level, the theories (for both signs of the mass)
are related by level-rank duality \cite{Naculich:1990pa,Camperi:1990dk,Mlawer:1990uv,Nakanishi:1990hj,Naculich:2007nc}.
This flow can be used to fix the finite shift in the Chern-Simons
level.

The second piece of evidence is that one can flow from supersymmetric dualities,
for which there is a lot of evidence at finite $N$, to the non-supersymmetric
dualities  \cite{Jain:2013gza,Gur-Ari:2015pca}. This flow is under precise control at large $N$, but the same
flow works also at finite $N$, at least as long as the finite-$N$ corrections
do not lead to qualitative changes in the flow (such as new operators becoming
relevant). 

In principle, both methods should enable us to determine whether the duality
relates $SU(N)$ theories or $U(N)$ theories, and also to determine the
precise level of the $U(1)$ factor in $U(N)$ (which also does not affect
the large $N$ tests). However, this was not done in the previous literature
on the subject. In section \ref{background} we use this method to fix the precise form of
the duality. We find that there is one version of the duality which relates
$SU(N)$ to $U({\tilde N})$ theories, and another that relates
$U(N)$ to $U({\tilde N})$ theories, but where the level of the $U(1)$
is not equal to the level of $SU(N)$.

The subtle distinctions above do not affect much of the physics of these
theories, but it greatly modifies the form of operators whose dimension
scales as a power of $N$ in the large $N$ limit. $SU(N)$ CS-matter theories contain baryon operators charged under
a $U(1)$ global symmetry, while
$U(N)$ theories do not. $U(N)$ theories contain monopole operators
charged under a $U(1)$ global symmetry,
and the form and dimension of the gauge-invariant globally-charged monopole operators
depends strongly on the level of the $U(1)$ group (and not just the
$SU(N)$ group). Thus, we can learn about the precise form of the duality
by matching these ``non-perturbative'' (in the sense of the $1/N$
expansion) operators between the two sides. 

A naive computation of the monopole dimensions on the two sides of
the duality of \cite{Giombi:2011kc,Aharony:2011jz,Aharony:2012nh} leads to a serious mismatch, apparently contradicting the
duality \cite{Pufu}. It was suggested in \cite{Radicevic:2015yla} that this
mismatch could be partly resolved if the duality relates $SU(N)$
theories to $U({\tilde N})$ theories. Our discussion, motivated by
 \cite{Radicevic:2015yla}, shows that
this is indeed the case, and we will show that this actually completely
resolves the mismatch between the two sides.

In section \ref{bmcs} we review
the properties of the baryon and monopole operators, in CS-fermion
and CS-scalar theories. In some cases the dimensions of these operators
in the large $N$ 't Hooft limit are known, and in other cases they are
only known at weak coupling. Some of these operator dimensions turn
out to scale as $N$ in the 't Hooft limit, and others as $N^{3/2}$.

In section \ref{bmduality} we show that the baryon and monopole operators precisely
map to each other under the duality in the $SU(N) \leftrightarrow
U({\tilde N})$ case, and that the monopole operators precisely map
to each other in the $U(N) \leftrightarrow
U({\tilde N})$ case. More precisely, we cannot really show this since
we do not know how to compute the dimensions at all values of the
coupling. But we find that the quantum numbers of these operators
are the same on both sides, and that the weak coupling limit of the dimension in
one theory maps precisely to the weak coupling limit in the other
theory, suggesting that the large $N$ value of this dimension is
not corrected as a function of the 't Hooft coupling. This was shown for a specific operator (the
monopole in CS-fermion theories) in \cite{Radicevic:2015yla}, but the duality suggests
that it should be correct also for the other operators. 

In section \ref{highspin} we discuss how these non-perturbative operators should
appear in (the UV completion of) the high-spin gravity theories on $AdS_4$
that are dual to these Chern-Simons matter theories. These theories
have a $U(1)$ gauge field on $AdS_4$ that couples to the baryon number or to
the monopole number; we argue that these two charges are related
by electric-magnetic duality. We argue that the high-spin gravity
theories should contain (in some $SL(2,Z)$ frame of the electric-magnetic
duality of this $U(1)$ gauge field) magnetically-charged particles with an energy scaling as
$N$ (so that they could be classical solutions of the Vasiliev action),
and electrically-charged particles with an energy scaling as
$N^{3/2}$. The energy of these particles (in Planck units)
should depend strongly
on the theta angle appearing in the Vasiliev equations of motion.
It would be interesting to understand the implications of these
particles for the non-perturbative completion of the high-spin
gravity theories.

\section{Dualities between Chern-Simons theories}
\label{background}

In this section we review the form of level-rank dualities between Chern-Simons theories,
and clarify their relation to dualities between Chern-Simons-matter theories. We discuss
here only $SU(N)$ and $U(N)$ theories.

The value of the Chern-Simons level $k$ of an $SU(N)$ gauge theory depends (as
many coupling constants do) on the regularization procedure. There are two
common regularizations that are used for Chern-Simons(-matter) theories. One
involves starting from a Yang-Mills-Chern-Simons theory at high energies, and
flowing to low energies where the Yang-Mills term becomes irrelevant. In most
diagrams gluon
loops are cut off by the fact that the $3d$ Yang-Mills theory is super-renormalizable.
In this regularization, $k$ is quantized (we normalize it so that it is quantized to
be an integer in the absence of matter fields), and can take any integer value.

Another common regularization of Chern-Simons(-matter) theories is dimensional 
regularization. This is somewhat subtle since the Chern-Simons interaction
involves an epsilon-symbol, but there is a standard way to perform it \cite{Chen:1992ee}.
The level (coupling constant) ${\hat k}$ appearing in this regularization is shifted compared
to the previous one, such that for any $k \neq 0$, 
\be \label{shift}
{\hat k} = k + N {\rm sign}(k).
\ee
In particular, $|{\hat k}| > N$. This shift arises due to a one-loop diagram with
a gluon loop. It is not present for $U(1)$ gauge groups.

When discussing dualities, it is important to be consistent about the regularization.
The Chern-Simons level-rank duality is usually discussed in the Yang-Mills regularization.
When we have a $U(N) = (SU(N)\times U(1)) / Z_N$ theory, we can apriori have different 
levels for $SU(N)$ and $U(1)$ (consistent with the global structure of the gauge group).
We will denote $U(N)$ CS theories as $U(N)_{k_1,k_2}$ where $k_1$ is the $SU(N)$
level (quantized to be an integer), and $k_2$ is the $U(1)$ level (normalized so that
$U(N)_{k,k}$ is the theory one gets by a Chern-Simons term involving a trace in
the fundamental representation of $U(N)$; note that other normalizations are
often used in the literature). In this convention, Chern-Simons theories are
invariant under a level-rank duality relating\footnote{Here and in some places below
we assume without loss of generality that $k>0$. The results for $k<0$ follow by
a parity transformation.} \cite{Naculich:1990pa,Camperi:1990dk,Mlawer:1990uv,Nakanishi:1990hj,Naculich:2007nc}
\be
SU(N)_k \leftrightarrow U(k)_{-N,-N},
\ee
and another useful level-rank duality relates
\be
U(N)_{k,k+N} \leftrightarrow U(k)_{-N,-k-N}.
\ee
In particular, the $SU(N)$ theory maps to the simplest $U(k)$ theory, that has equal levels
for its two factors.

Using \eqref{shift} (and the fact that there is no shift of $U(1)$ levels), it is easy to 
express the same dualities in the dimensional
regularization conventions. To avoid confusion we use hatted indices for these
conventions, and then we have
\be \label{dr1}
SU({\hat N})_{\hat k} \leftrightarrow U({\hat k}-{\hat N})_{-{\hat k}, -{\hat N}}
\ee
and
\be \label{dr2}
U({\hat N})_{{\hat k},{\hat k}} \leftrightarrow U({\hat k}-{\hat N})_{-{\hat k},-{\hat k}}.
\ee
In this convention the second level-rank duality relates the simplest $U({\hat N})$
theories.

Consider now the duality of \cite{Giombi:2011kc,Aharony:2011jz,Aharony:2012nh} relating $U(N)$ Chern-Simons theories coupled to $N_f$
fermions in the fundamental representation, to Chern-Simons theories coupled to $N_f$ scalars in the
fundamental representation (the duality also involves
a Legendre transform with respect to the singlet scalar operator in one of the theories, but this is not
important for any of the issues we discuss in this note). In the large $N$ limit, the 't Hooft coupling in the dimensional
regularization procedure is defined by $\lambda \equiv {\hat N} / {\hat k}$, and in the Yang-Mills
regularization by $\lambda \equiv N / (k+N {\rm sign}(k))$. In both cases $-1 < \lambda < 1$, and in
the large $N$ limit it is effectively a continuous variable. Various quantities were
computed exactly as a function of this 't Hooft coupling in the large $N$ limit of these non-supersymmetric field
theories
\cite{Banerjee:2012gh,Aharony:2012nh,GurAri:2012is,Radicevic:2012in,Aharony:2012ns,Jain:2013py,Takimi:2013zca,Jain:2013gza,Frishman:2013dvg,Bardeen:2014paa,Jain:2014nza,Frishman:2014cma,Moshe:2014bja,Bedhotiya:2015uga,Gur-Ari:2015pca,Radicevic:2015yla,Geracie:2015drf}, and were
all found to be consistent with the duality.

In \cite{Aharony:2012nh} it was argued that the precise form of this
duality for finite $N$ could be determined by deforming the fermionic theory by a fermion
mass such that it flows at low energies to a pure CS theory (depending on the sign of
the mass), and mapping that deformation
to the scalar side where one obtains a different pure CS theory (that theory again depends
on the sign of the mass, because for one sign there is a scalar condensate partly
breaking the gauge group). The two low-energy CS theories
must then be related by level-rank duality. The arguments of \cite{Aharony:2012nh} were not careful about the
precise form of the level-rank duality, and about whether we have an $SU(N)$ or $U(N)$
gauge group, and they found that (in the Yang-Mills regularization) the duality should 
relate the theory of $N_f$ fundamental scalars coupled to $U(N)_k$, to a theory of
$N_f$ fundamental fermions coupled to $U(k)_{N_f/2-N}$. Using the precise form
of the level-rank dualities above, we see that there are actually three possible dualities
that are consistent with level-rank duality:
\be\label{dual1}
SU(N)_k {\rm\ coupled\ to\ scalars} \leftrightarrow U(k)_{-N+N_f/2,-N+N_f/2} {\rm \ coupled\ to\ fermions},
\ee
\be\label{dual2}
U(N)_{k,k} {\rm\ coupled\ to\ scalars} \leftrightarrow SU(k)_{-N+N_f/2} {\rm \ coupled\ to\ fermions},
\ee
\be\label{dual3}
U(N)_{k,k+N} {\rm\ coupled\ to\ scalars} \leftrightarrow U(k)_{-N+N_f/2,-N-k+N_f/2} {\rm \ coupled\ to\ fermions}.
\ee
The evidence so far for the validity of these dualities is identical, since it is just
at leading order in the large $N$ limit, where the differences between them are
negligible. We conjecture
that they are all correct. The main difference between the different 
theories is in the
baryonic and monopole operators, and we will discuss these operators and their
consistency with the duality in the next sections.

It was argued in \cite{Aharony:2012ns} that when the $U(N)$ Chern-Simons-matter theories are compactified on a
circle, the logarithms of the eigenvalues of their holonomies are quantized to be an integer multiple
of $2\pi / {\hat k}$. In the $SU(N)$ theory the eigenvalues sum up to zero (up to
an overall shift by a multiple of $2\pi$), and the quantization applies to the
difference between any pair of eigenvalues. This quantization has important
effects for the thermodynamical behavior of these theories, which was analyzed
in the large $N$ limit in \cite{Aharony:2012ns,Jain:2013py} (following \cite{Giombi:2011kc}).

Note that the $SU(N)$ theory has a $U(1)$ baryon-number symmetry, carried by the
fields in the fundamental representation.
The $U(N)$ theories also have a $U(1)$ global symmetry, which is a
combination of the baryon-number symmetry and the ``topological''
symmetry generated by the current $J^{\mu} = \epsilon^{\mu \nu \rho}
{\rm tr}(F_{\nu \rho})$. One combination of these two currents is gauged,
and any other combination is a global symmetry, that maps under the duality
to the corresponding global symmetry on the other side. The precise
combination that is gauged depends on the Chern-Simons level of the $U(1)$ 
gauge field; it is independent of the $SU(N)$ Chern-Simons level.

Let us end this section by clarifying the relation of these dualities to dualities
between ${\cal N}=2$ supersymmetric CS-matter theories 
\cite{Giveon:2008zn,Benini:2011mf,Aharony:2013dha,Aharony:2014uya}.
For these theories the Yang-Mills regularization, which preserves
supersymmetry, is usually used. However, the supersymmetric pure Yang-Mills-Chern-Simons
theory contains two extra massive gluinos in addition to the pure CS
theory. Integrating these out gives a shift between the level of the ${\cal N}=2$
supersymmetric $SU(N)$ CS theory and the non-supersymmetric one, of the form
\be \label{shift2}
k^{{\cal N}=2} = k^{{\cal N}=0} + N {\rm sign}(k^{{\cal N}=0}).
\ee
This is precisely of the same form as \eqref{shift}, though their origin is
completely different (a loop involving gluons in one case, a loop with two
gluinos in the other). Thus, the level-rank dualities for pure ${\cal N}=2$
CS theories take the same form as \eqref{dr1} and \eqref{dr2} above.
This is consistent with the form of the known supersymmetric dualities
between Chern-Simons-matter theories with chiral multiplets in the
fundamental and/or anti-fundamental representation \cite{Giveon:2008zn,Benini:2011mf,Aharony:2013dha,Aharony:2014uya};
in fact taking $N_f \to 0$ in the equation for the mapping between the
two sides of these dualities gives
precisely the mappings \eqref{dr1} and \eqref{dr2}. When we flow
from ${\cal N}=2$ supersymmetric
Yang-Mills-Chern-Simons-matter dualities to non-supersymmetric CS-matter dualities, as
described in \cite{Jain:2013gza,Gur-Ari:2015pca}, both for the $SU(N) \leftrightarrow U({\tilde N})$
dualities and for the $U(N) \leftrightarrow U({\tilde N})$ dualities, we recover \eqref{dual1}-\eqref{dual3},
taking into account the
shift in the Chern-Simons level arising from integrating out
massive fermions. Note that apriori ${\cal N}=2$ Chern-Simons-matter
theories exist for any value of $k^{{\cal N}=2}$, but \eqref{shift2}
suggests that if we try to flow from there to ${\cal N}=2$
supersymmetric pure Chern-Simons theories by giving masses to all
the flavors, we will only obtain
theories with $|k^{{\cal N}=2}| > N$ (in other cases this flow
leads to supersymmetry breaking). This is consistent with
all cases we are familiar with.

\section{Baryons and monopoles in Chern-Simons-matter theories}\label{bmcs}

In the large $N$ limit, the finite-dimension operators in all the
theories discussed in the previous section are the same. They
are just the bi-fundamental operators, schematically
$\Phi^{\dagger} D D \cdots D \Phi$ or ${\bar \psi} D D \cdots
D \psi$ (these include currents of all spins, that are conserved
in the large $N$ limit for any value of $\lambda$), and their ``multi-trace'' products. However, operators
whose dimensions scale with $N$ are significantly different in
the different theories discussed above, and we review their
properties in this section.

Let us begin with the $SU(N)$ theories, and for simplicity let us take a
single matter field in the fundamental representation, $N_f=1$ (the
generalization to arbitrary values of $N_f$ is straightforward; for 
$N_f=1$ the CS level in CS-fermion theories must be a half-integer). In such
theories we can form baryon operators by taking a product of $N$
fields in the fundamental representation and anti-symmetrizing
their color indices. We normalize the $U(1)$ global symmetry such
that the baryon operator carries charge $1$.

If we have a fermion in the fundamental
representation, we can just take a product of its $N$ color
components and anti-symmetrize them, and this gives the
lowest-dimension operator carrying baryon-number charge. This operator
has a classical dimension $\Delta = N$; of course this dimension can
obtain quantum corrections. Note that the fermions have two
different spin states, and each fermion can be independently
chosen to be in each of these states, so we actually have
$2^N$ baryon operators of this type, of various spins ranging
up to $N/2$.

For scalars in the fundamental
representation we cannot do this since the $N$ bosonic
operators that we multiply must all be different in order for
the baryon operator to be non-zero. So to form
baryon operators we need to take some derivative operator
acting on each color component of the scalar, with a different
derivative operator for each component. Given that the
scalar obeys (in the free theory) $\partial^2 \phi = 0$, the
number of different operators with $j$ derivatives acting
on the scalar is $2j+1$. Choosing the minimal number
of derivatives that are all different, we find that the minimal dimension
baryonic operator that we can form from scalars has
a dimension that scales as $\Delta \simeq \frac{2}{3} N^{3/2}$ for large $N$
\cite{Shenker:2011zf}.
Again there are many operators of this type, of different
spins, depending on precisely how we contract the
different derivatives. 

In $U(N)$ theories baryons are not gauge-invariant, but a similar
role is played by monopole operators. These operators are
characterized by a gauge flux around them, which can be chosen
to be in the Cartan subalgebra of the gauge group. When the
gauge group is $U(N)$ (rather than $SU(N)\times U(1)$), the
monopole operators are characterized by $N$ integers $q_i$ (up
to permutations in the Weyl group), which are the magnetic
fluxes in $U(1)^N \subset U(N)$ (these are known as the GNO charges). 
We normalize the global ``topological'' $U(1)$
charge so that this monopole operator carries a global charge $\sum_{i=1}^N q_i$.
Note that monopole operators with different magnetic fluxes
that carry the same global charge can mix with each other in
the quantum theory.

In a Yang-Mills theory these monopoles are gauge-invariant operators,
but in a (Yang-Mills-)Chern-Simons theory the Chern-Simons coupling 
gives the monopole operators an electric charge. The simplest possibility
to analyze is the $U(N)_{k,k}$ theory. In this theory a generic monopole
operator, with $n$ different non-zero charges $q_i$, breaks the
gauge group to $U(1)^n\times U(N-n)$, and carries a charge
$q_i k$ under the $U(1)_i$ unbroken gauge group. This charge must
be balanced by an appropriate number of fields in the (anti-)fundamental 
representation of $U(N)$ in order to obtain a gauge-invariant operator.  Note that
$k$ here is defined in the Yang-Mills regularization, since the monopoles
are naturally defined in the high-energy Yang-Mills-Chern-Simons
theory.

Let us consider for simplicity the monopole operators with charges
$\{q_i\} = (1, 0, \cdots, 0)$. These are expected to be the
lowest-dimension operators carrying the global $U(1)$ charge.
In this case we need to multiply the monopole by $|k|$ fields
in the fundamental or anti-fundamental representation, which all
carry the same gauge group index. In the scalar theory we can
just put in $|k|$ scalars, so the dimension of the corresponding
operator is classically $\Delta = |k| \Delta_{scalar}+\Delta_{monopole}$, where
$\Delta_{monopole}$ is the dimension of the monopole operator
itself (this can be thought of as the difference in energy between
the lowest state with flux $\{q_i\} = (1, 0, \cdots, 0)$ on $S^2$, before
adding the necessary fields to make it gauge-invariant,
and the lowest zero-flux state). This difference vanishes in the non-supersymmetric
CS-matter theory with no fluxes for background fields; in theories with
background fluxes (which is usually the case in supersymmetric
theories) it can be non-zero. $\Delta_{scalar}$ here is the
dimension of the scalar in a monopole background, which is the
same as the energy of the lightest scalar state on $S^2$ in
the flux background\footnote{The discussion here is not completely
rigorous, since the monopole separately and the scalar separately are
not gauge-invariant. However, it is valid at least at weak coupling
(and in supersymmetric theories for chiral operators), which is enough
for our purposes here.}. For unit flux it is $\Delta_{scalar}=1$;
note that this is shifted from the lowest energy without the
flux, that gives $\Delta_{scalar}=\frac{1}{2}$.
More over, this scalar state carries spin one half \cite{Wu:1976ge}, such
that we have $2^{|k|}$ states of this type, with $\Delta=|k|$ (classically),
carrying different
spins up to $|k|/2$.

In the fermionic theory, as in our discussion of the baryons above,
we need to choose $|k|$ different fermionic operators multiplying
the monopole in order for the resulting gauge-invariant operator
to be non-zero. Thus, as in our analysis above, the classical
dimension of the monopole operator in the large $N$ limit is $\Delta \simeq \frac{2}{3} |k|^{3/2}$. 
Note that there is also a contribution
here from the fermion operators (involving the dimension of the
fermion operators in the monopole background), but this
is negligible compared to the contribution from the
derivatives. The lowest fermion state in the monopole background
carries spin zero \cite{Wu:1976ge}, but the gauge-invariant monopole operators we obtain can have
many different spins, depending on precisely how we
contract the indices on the derivatives.

If we have different levels for the $SU(N)$ and the $U(1)$ the
analysis changes. Let us first discuss the ``bare'' monopole operators
for $U(N)_{k,k'}$ theories, assuming $k,k' > 0$. It is convenient to decompose
the charge of the lowest monopole into its $U(1)$ and $SU(N)$
components :
\be
\{q_i\} = (1,0,\cdots,0) = \left({1\over N},{1\over N},\cdots,{1\over N}\right) +
\left(\frac{N-1}{N}, -{1\over N},\cdots,-{1\over N}\right).
\ee
When the $U(1)$ has CS level $k'$, the $U(1)$ electric charge of
this monopole is equal to $k'$ times that of the fundamental
representation. When the $SU(N)$ has CS level $k$,
the $SU(N)$ electric charge is in the symmetric product of $k$
fundamental representations of $SU(N)$. Thus, when the levels
are equal, this charge is the same as that of $k$ fields in the
fundamental representation of $U(N)$, and the monopole can
be made neutral by multiplying it by $k$ fields in the
anti-fundamental representation, as above.

When $k'$ is not equal to $k$, the electric charge we obtain
is only a consistent charge in $U(N)$ when $k-k'=0$ (mod $N$),
a constraint that is presumably required for gauge-invariance of
the $U(N)$ Chern-Simons term\footnote{Here it is important that
the gauge group is precisely $U(N) = (SU(N)\times U(1)) / Z_N$; we will
not discuss other quantizations of the $U(1)$ charge here.}. If
(say) $k'=k+N$ (the case that appears in \eqref{dual3})
then the monopole carries $N$ extra units of $U(1)$ charge
 in the fundamental representation compared to the product of $k$ fields in the fundamental
representation of $U(N)$. Thus its charges are the same as an $SU(N)$ baryon
added to that product (and it can be made neutral by
multiplying it by an anti-baryon and $k$ fields in the
anti-fundamental representation). The fundamental fields
that we need to add to make the monopole gauge-invariant
obey similar symmetry constraints to the ones above;
similar to monopoles for the fields with color index one,
and similar to baryons for the fields with other color indices.
Thus, in the scalar theory we obtain by the same analysis as
above a classical dimension of order $\frac{2}{3} N^{3/2}$
at large $N$, and in the fermionic theory we obtain a
classical dimension of order $\frac{2}{3} |k|^{3/2}$ for
this minimal globally-charged operator in the $U(N)_{k,k+N}$
theory. The generalization to other values of $(k'-k)$
is straightforward; if $k'=k+n N$ for some integer $n$,
the lowest-dimension operator carrying the
minimal global $U(1)$ charge would look like
a bound state of a monopole and $n$ baryons.

\section{Mapping of baryons and monopoles under the duality}\label{bmduality}

Dualities between different Chern-Simons-matter
theories should map the lowest-dimension operator with some spin and
global charge to the lowest-dimension operator with the
same spin and global charge in the dual theory. For dualities
between $SU(N)$ and $U({\tilde N})$ theories, this means
that baryons should be mapped to monopoles, and this is indeed
what happens for chiral monopole and baryon operators 
in supersymmetric dualities of this type \cite{Aharony:2013dha,Aharony:2014uya}.

In the non-supersymmetric case, let us start from the duality
\eqref{dual2}. In the fermionic $SU(k)_{-N+1/2}$ theory the globally
charged operators are baryons, with a lowest dimension $|k|$ at
weak coupling. In the bosonic $U(N)_{k,k}$ theory, the
globally charged operators are monopoles, and their
lowest classical dimension is $|k|$. Thus, there is an exact
match between the classical dimensions on both sides
(even at finite $N$ and $k$). This suggests that these
dimensions do not acquire quantum corrections (at large
$N$, and maybe even at finite $N$), even though we
do not know any reason for this. Note that on both
sides we have $2^{|k|}$ operators of different spins,
in agreement with the duality. For $N_f > 1$ one can
check that the global $SU(N_f)$ quantum numbers of the
baryons and monopoles also match between the two sides.

In the duality \eqref{dual1}, the lightest globally-charged 
operators in the bosonic $SU(N)_k$ theory are baryons,
whose dimension at weak coupling scales as $\Delta \simeq \frac{2}{3} N^{3/2}$.
In the fermionic $U(k)_{-N+1/2,-N+1/2}$, the lowest-dimension
charged operators are monopoles, whose dimension in the large
$N$ limit scales as $\Delta \simeq \frac{2}{3} N^{3/2}$.
So again we find a precise match of the classical dimensions
in the large $N$ limit. In
both cases the degeneracy and spins arise from the different
ways of contracting derivatives acting on spin zero operators,
so these also match between the two sides, as do the
$SU(N_f)$ representations when $N_f > 1$.

It was argued in \cite{Radicevic:2015yla} that the monopole dimension in the
fermionic theory at large $N$ is independent of the 't
Hooft coupling $\lambda$, and thus that the result above
$\Delta \simeq \frac{2}{3} N^{3/2}$ is actually valid for
any value of $\lambda$. The reasons for this absence of
quantum corrections are not completely clear (it was
argued in \cite{Aharony:2015pla} that general diagrams would contribute
corrections to $\Delta$ that could even change its
large $N$ scaling), but it
follows from the thermal partition function computation of \cite{Radicevic:2015yla}.
The duality then suggests that also the dimensions of the
baryons in CS-scalar theories do not receive quantum corrections in the large
$N$ limit; again there is no obvious reason for this to
be true. 

Note that the analysis of \cite{Radicevic:2015yla}
was naively done using the dimensional regularization convention, 
with the eigenvalues of the holonomies of the $U(N)_{k,k}$ theory
on a circle separated by distances $2\pi / |{\hat k}|$.
Here the relevant level is the $SU(N)$ level, so
the computation in \cite{Radicevic:2015yla} was done for an 
$SU(N)$ level $|k| = |{\hat k}| - N$.
Dimensional regularization was indeed used in \cite{Radicevic:2015yla} (following
\cite{Aharony:2012ns,Jain:2013py}) to define the
level in the computation of the eigenvalue distribution
(which here involves the $(N-1)$ eigenvalues that do not have any background
monopole flux). However, the contribution to the dimension
of the monopole operator in \cite{Radicevic:2015yla} actually
comes from the integration
over the single eigenvalue corresponding to the $U(1)$
subgroup in which the monopole flux was turned on. This
integration was performed using the fact that the monopole
background with charges $\{q_i\}$ induces $k q_i$ units of background 
charge for the $i$'th $U(1)$ gauge group. As we discussed
above, this is true for the $U(N)_{k,k}$ theory where
$k$ is defined using the Yang-Mills regularization. In the
analysis of \cite{Radicevic:2015yla} the result only depends
on
$k$, going as (for the simplest monopole) $\Delta \simeq \frac{2}{3} |k|^{3/2}$ 
(independently of $N$). Thus, this analysis should be interpreted
as being done in the $U(N)_{k,k}$ theory (whose dimensional
regularization $SU(N)$ CS level is ${\hat k} = k + N {\rm sign}(k)$). Therefore,
our statement in the previous paragraph about the
implications of the analysis of \cite{Radicevic:2015yla} is
indeed correct in the fermionic $U(k)_{-N+1/2,-N+1/2}$
theory that we mentioned there.

In the duality \eqref{dual3}, we saw in the previous section  that the classical dimension
of the lightest globally-charged state is the sum of the dimensions
for baryons and monopoles. So by joining together our analysis
of the two previous cases, these lightest states map to each other also
under this duality \eqref{dual3}, again assuming that the dimensions of the
corresponding operators have no corrections in the large $N$ limit. 

In the supersymmetric Chern-Simons-matter dualities, we can
explicitly map the chiral baryon or monopole operators, since
we can read off their dimensions from the supersymmetric index
\cite{Romelsberger:2005eg,Kinney:2005ej,Bhattacharya:2008zy}. This mapping was not yet performed for $SU(N)$ dualities,
but it was performed for some dualities between $U(N)$ CS-matter theories
in \cite{Aharony:2015pla}. Obviously, since the partition functions are the same on
both sides, the monopole operators were also found to match.
However, surprisingly, it was found that the dimensions of the
lowest chiral monopole operators are often much larger than
the dimensions computed above (the relevant dimensions are the ones
for theories with scalar
fields, since such fields are present in the supersymmetric theories), and can
even scale differently with $N$. This
implies that in these theories the dimensions of the monopole
operators that classically have the lowest dimensions are usually not
protected. The analysis above suggests that the anomalous
dimensions of at least some of these operators are negligible, at least in the large $N$
limit, but it is not known how to test this directly.

\section{Mapping to high-spin gravities}\label{highspin}

The $U(N)_k$ Chern-Simons-matter theories discussed above are
conjectured \cite{Giombi:2011kc,Aharony:2011jz,Chang:2012kt,Aharony:2012nh} (following the $k=0$ case discussed in \cite{Klebanov:2002ja,Sezgin:2003pt})
to correspond to Vasiliev's high-spin gravity theories
on $AdS_4$. For $N_f=1$ they are mapped to the original
Vasiliev theory, while for higher $N_f$ they map to a
version of this theory where all the fields transform in the
adjoint representation of $U(N_f)$. The coupling constant of the Vasiliev theories in
this mapping goes as $1/N$ (see a more detailed discussion below), 
and since these theories are only
understood classically, this duality is well-defined only in the
large $N$ limit; the Chern-Simons-matter theories can be viewed as giving
a non-perturbative completion of the Vasiliev theories. In a
special supersymmetric case the string theory completion of
the Vasiliev theory is also known (at least formally) \cite{Chang:2012kt}, so this
string theory can also be viewed as a UV completion (identical
to the one provided by the CS-matter theory).

Since the duality is only understood at large $N$, it is not
clear which, if any, of the CS-matter theories discussed in the previous sections corresponds
to the Vasiliev theory. In fact, they all do. This is because
we can move between the different CS-matter theories by shifting the $U(1)$
level and gauging or ``ungauging'' the $U(1)$, and in the
context of the Vasiliev theory, as discussed extensively in
\cite{Chang:2012kt} (following \cite{Witten:2003ya}), these transformations just
involve a modification of the boundary conditions of the $4d$
$U(1)$ gauge field of the Vasiliev theory. Note that this is
consistent with the fact that the operators in these theories
whose dimension does not scale with $N$ are identical.

In particular, let us begin with the Vasiliev
theory corresponding to the $SU(N)_k$ theory coupled
to scalars (appearing in \eqref{dual1}). This theory (up to the $U(1)$ issues) is
believed to be dual to the Vasiliev theory with a
theta angle related to the $SU(N)$ 't Hooft coupling
by $\theta = \frac{\pi}{2} \lambda$ \cite{Chang:2012kt,Aharony:2012nh,Giombi:2012ms}.  This
Vasiliev theory includes a $4d$ $U(1)$ gauge field, which is related in the usual 
way to the global baryon number symmetry in the $3d$ $SU(N)$
CS-matter theory; thus the charged states in the
bulk should include the baryons of this theory. As
discussed above, the same theory should have a
description as a $U(k)_{-N,-N}$ theory coupled to
fermions, and in that description the same charged
states are monopoles of the CS-matter theory.

We can take this theory and modify the boundary
conditions on the $4d$ $U(1)$ gauge field, in a way
that corresponds to gauging the $U(1)$ global
symmetry of the $SU(N)$ CS-matter theory. This involves
making the boundary value of the $4d$ $U(1)$ gauge
field dynamical, instead of the subleading term
in the expansion of this gauge field near the
boundary \cite{Witten:2003ya}; one should also be careful about the charge-quantization conditions to 
ensure that the resulting theory has gauge group $U(N)$ rather than $SU(N)\times U(1)$.
 We can then add for this $3d$ gauged $U(1)$ a CS term of
level $k'$ (which has no relation to $k$). For
specific choices of $k'$ this gives the theories
appearing in \eqref{dual2} and \eqref{dual3}
above, and our discussion above implies that
the interpretation of this procedure on the
fermionic side is as changing the theory on the
right-hand side of \eqref{dual1} to that on the
right-hand side of \eqref{dual2} or \eqref{dual3}.
Note that if we start from a $U(N)$ gauge theory
and gauge its global ``topological'' $U(1)$ symmetry, we generally
get a $U(N)\times U(1)$ theory, with a mixed
CS coupling between the $U(1) \subset U(N)$ and the new $U(1)$.
However, in some cases this turns out to be
equivalent to an $SU(N)$ theory (see, for
instance, the discussion in \cite{Aharony:2013dha}; this is called
``ungauging''), and presumably
this also happens here in the fermionic theory when
we go from \eqref{dual1} or \eqref{dual3} to
\eqref{dual2}.

As discussed in \cite{Witten:2003ya}, the operation of gauging a
$3d$ $U(1)$ global symmetry maps by the AdS/CFT correspondence to
an electric-magnetic
duality on the $4d$ bulk gauge
field, and the operation of shifting the
Chern-Simons level of this $3d$ gauged $U(1)$ is the
same as shifting the bulk theta angle (multiplying $F\wedge F$)\footnote{Here we
refer to the theta angle of the $U(1)$ gauge theory on $AdS_4$. This should not be
confused with the theta angle mentioned elsewhere in this section, that appears
in the equations of motion of the Vasiliev theory.} by a multiple of $2\pi$ (this has no
effect on the bulk physics, but does affect
the physics in the presence of the boundary).
So all the operations described in the
previous paragraph can be interpreted as
$SL(2,Z)$ transformations on the bulk $U(1)$
gauge theory. In particular, if in the $SU(N)$
picture the allowed charges of states in the bulk were
electric charges of the bulk $U(1)$ (with the
total $4d$ magnetic charge vanishing because of
the boundary conditions), after these
operations the only allowed charges are dyonic
(with the ratio of the electric and magnetic
charge depending on $k'$).

As we saw in the previous section, this modification
of the allowed charges is mirrored in the gauge
theory by the fact that instead of having baryon
operators we now have monopole operators
(that also carry some baryon number charge).
The monopole charge of the CS-matter theory
should thus be viewed as the electric-magnetic
dual in the bulk of the baryon number charge.

At first sight we have a contradiction, since on
one hand the only difference between the theories
\eqref{dual1}, \eqref{dual2} and \eqref{dual3}
on the gravity side is in the boundary conditions,
but on the other hand their spectrum of
operators is completely different. However, 
the bulk theory in all these cases is actually
exactly the same, and it should thus contain
excitations corresponding to all the baryon
and monopole operators that we discussed
above. From the bulk point of view, the reason
for the different spectrum of these theories is
that the different boundary conditions result
in different Gauss' law constraints on the
total charge. The constraint for the $SU(N)$
theory allows only non-zero total electric (=baryon number)
charge, while the one of $U(N)$ theories
allows only a specific dyonic charge
(which we called monopole number).
So, while there are bulk excitations corresponding
to both baryons and monopoles, only a subset
of the multi-baryon and multi-monopole states,
which obeys Gauss' law, gives consistent
states in the full theory. The allowed charge in
all cases maps to the global $U(1)$ charge of the
$3d$ Chern-Simons-matter theory.

If we interpret the Vasiliev theory from the point
of view of the scalar theory, then the baryons
have a dimension going as $N^{3/2}$, while the
monopoles have a dimension scaling as $|k|$; from
the fermionic point of view the same numbers are $|k|^{3/2}$
and $N$. 

The action for the Vasiliev theory is not yet
known, but only its equations of motion, which determine
the action up to an overall scaling. This scaling may
be determined, for instance, from the two-point
function of the energy-momentum tensor in the
CS-matter theory, which should be related to the
Einstein term in the bulk action (it is not affected by the theta parameter).
This two-point function is proportional in the scalar
interpretation to $N \frac{\sin(\pi \lambda)}{\pi \lambda}$
 \cite{Aharony:2012nh}, which for any $-1 < \lambda < 1$ in the
't Hooft limit scales as $N$. 

Thus, from the point
of view of the scalars, the dimension of the monopole
operators $\Delta \simeq |k|$ is consistent with the monopoles being classical
solutions of the Vasiliev theory in the bulk or of its UV completion, which carry
a magnetic $U(1)$ charge in our conventions above (though it does
not prove that such classical solutions exist). These solutions can be
thought of as charged black holes or charged solitons in the Vasiliev
theory. The discussion
above implies that the energy of these classical solutions,
in units set by the Vasilliev gravitational coupling, should
depend on the theta angle appearing in the Vasiliev
equations of motion. In particular it diverges (in these
units) in the $\theta \to 0$ limit. It would be very interesting to
find such classical solutions, if they exist, in order to
confirm this picture.

On the other hand, the baryons (from the scalar point of
view) with $\Delta \simeq N^{3/2}$ should be much heavier than this scale, and cannot be viewed as finite
excitations on top of the classical Vasiliev theory. So there should not
be any classical electrically-charged solutions.

The generalization of all our discussions to the supersymmetric
Chern-Simons-matter theories, whose gravitational duals were
discussed in \cite{Chang:2012kt}, seems to be straightforward, and we will
not perform it here. The main difference is that in these
theories, since they include both scalars and fermions in the
fundamental representation, the dimensions of the lowest-dimension
baryons and monopoles scale as $N$ and $|k|$, respectively, so both
of these may correspond to classical solutions. For
theories with ${\cal N} \geq 2$ supersymmetry, the
precise dimensions of the lowest-dimensional chiral baryon or
monopole operators may be found along the lines of \cite{Aharony:2015pla}.
These could correspond to classical solutions of the bulk theory that preserve
some of the supersymmetry.

\section*{Acknowledgements}

I would like to thank Guy Gur-Ari, Zohar Komargodski and Tarun Sharma for useful discussions, and
especially Djordje Radi\v cevi\' c for useful discussions and comments on a draft of this manuscript.
This work was supported in part by an Israel Science Foundation center for excellence grant (grant no. 1989/14), by the Minerva foundation with funding from the Federal German Ministry for Education and Research, by the I-CORE program of the Planning and Budgeting Committee and the Israel Science Foundation (grant number 1937/12), by the Henry Gutwirth Fund for Research, and by the ISF within the ISF-UGC joint research program framework (grant no. 1200/14). OA is the incumbent of the Samuel Sebba Professorial Chair of Pure and Applied Physics.

\bibliographystyle{JHEP}
\bibliography{baryon-monopole}

\end{document}